\documentclass[aps,prl,preprint,superscriptaddress,showpacs]{revtex4}
\usepackage{amsmath,epsfig}
\usepackage{graphicx}

\def\be{\begin{equation}}
\def\ee{\end{equation}}
\def\bea{\begin{eqnarray}}
\def\eea{\end{eqnarray}}

\def\bfr{{\bf r}}
\def\bfrp{{\bf {r^\prime}}}
\def\bfE{{\bf E}}
\def\bfH{{\bf H}}

\def\bfp0{{\bf{p_0}}}

\def\kjperp{k_{j\perp}}
\def\kp{k_{\parallel}}

\begin{document}
\title{Quantum Optics with Surface Plasmons}

\author{D.E.\ Chang}
\affiliation{Physics Department, Harvard University, Cambridge, MA 02138}

\author{A. S.\ S\o rensen}
\affiliation{Niels Bohr Institute, DK-2100 Copenhagen \O, Denmark}

\author{P.R.\ Hemmer}
\affiliation{Physics Department, Harvard University, Cambridge, MA 02138} 
\affiliation{Electrical Engineering Department, Texas A\&M University, College 
Station, TX 77843}

\author{M.D.\ Lukin}
\affiliation{Physics Department, Harvard University, Cambridge, MA 
02138}\affiliation{ITAMP, Harvard-Smithsonian Center for Astrophysics, 
Cambridge, MA 02138}

\date{\today}

\begin{abstract}
We describe a technique that enables strong, coherent coupling between 
individual optical emitters and guided plasmon excitations in conducting 
nano-structures at optical frequencies. We show that under realistic conditions, 
optical emission can be almost entirely directed into the plasmon modes.  As an 
example, we describe an application of this technique involving efficient 
generation of single photons on demand, in which the plasmon is efficiently 
out-coupled to a dielectric waveguide.
\end{abstract}

 \maketitle

The fields of quantum computation and quantum information science have spurred 
substantial interest in generating strong, coherent interactions between 
individual quantum emitters and photons. Such a mechanism would enable quantum 
information to be passed over long distances, which is essential for quantum 
communication~\cite{ekert91,briegel98} and would facilitate the scalability of 
quantum computers~\cite{svore04}.  The required coupling between emitters and 
photons is difficult but has been achieved in a number of systems that reach the 
so-called ``strong-coupling'' regime of cavity quantum 
electrodynamics~\cite{thompson92,brune96}. Recently, several approaches to reach 
this regime on a chip at microwave frequencies have been 
suggested~\cite{childress04,sorensen04,blais04,wallraff04} and experimentally 
observed~\cite{wallraff04}.  A key feature of these approaches is the use of 
conductors to reduce the effective mode volume $V_{\footnotesize\textrm{eff}}$ 
for the photons, thereby achieving a substantial increase in the coupling 
strength $g{\propto}1/\sqrt{V_{\footnotesize\textrm{eff}}}$.  Realization of 
analogous techniques with optical photons would open the door to many potential 
applications in quantum communication and in addition lead to smaller mode 
volumes and hence faster interaction times.

\begin{figure}[p]
\begin{center}
\includegraphics[width=7cm]{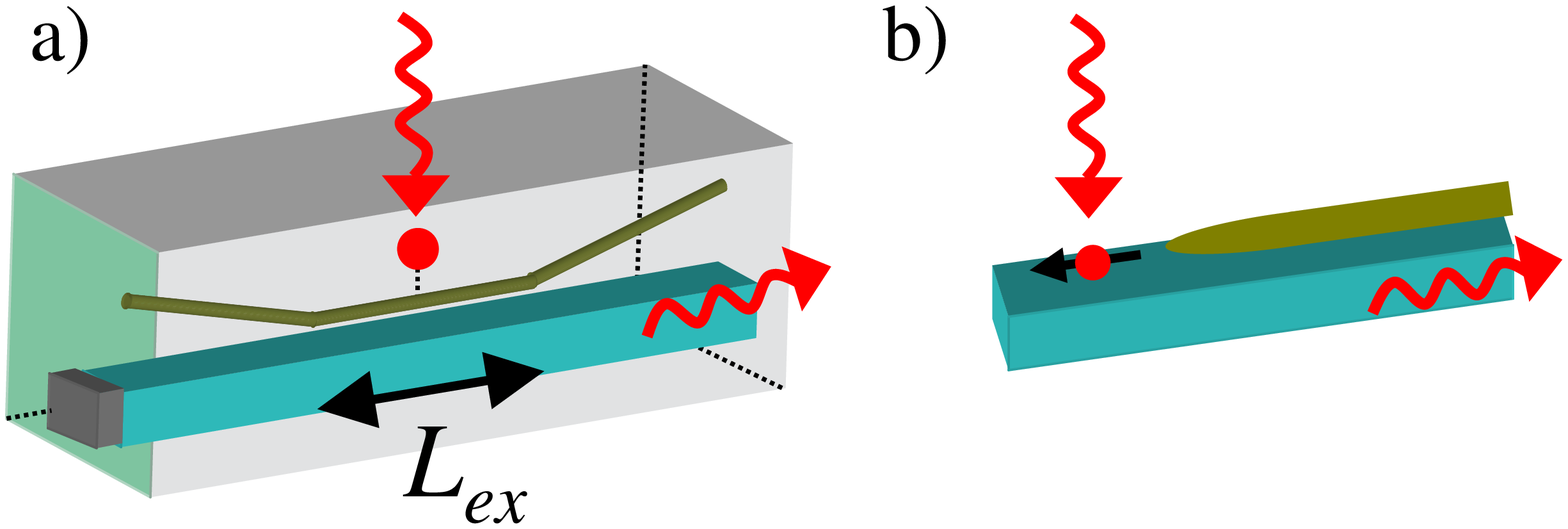}
\end{center}
\caption{(a) An emitter coupled to a nanowire is optically excited and decays 
with high probability into the plasmon modes of the nanowire.  A single photon 
source is created by evanescently coupling the nanowire to a nearby dielectric 
waveguide over a length $L_{ex}$. (b) A similar setup, where a dipole emitter is 
coupled to a metallic nanotip that expands to some final radius $R$ and is then 
coupled to a dielectric waveguide.\label{fig:applications}}
\end{figure}

This Letter describes a method that enables strong, coherent coupling between 
individual emitters and electromagnetic excitations in conducting 
nano-structures at optical frequencies, via excitation of guided plasmons 
localized to nanoscale dimensions.  The strong coupling is possible due to the 
small mode volume associated with this sub-wavelength confinement.  We first 
examine the simple case of a conducting nanowire, where the relevant physical 
mechanisms can be understood analytically, and then consider an optimized 
geometry of a metallic nanotip.  Because of dissipative losses in metals the 
plasmon modes themselves are not suitable as carriers of information over long 
distances.  We show, however, that the plasmon excitation can be efficiently 
out-coupled through evanescent coupling with a nearby dielectric waveguide, as 
illustrated schematically in Fig.~\ref{fig:applications}. This can be used, 
\textit{e.g.}, to create an efficient single photon source, or as part of an 
architecture to perform controlled interactions between distant qubits.  We find 
that single-photon efficiencies exceeding $95\%$ are possible using relatively 
simple implementations.

\begin{figure}[p]
\begin{center}
\includegraphics[width=9cm]{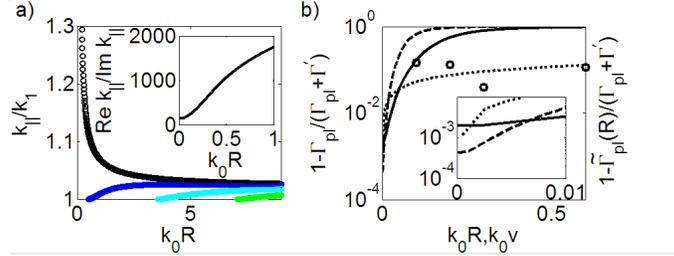}
\end{center}
\caption{(a) $\kp$ for plasmon modes on a silver nanowire as a function of wire 
radius $R$, in units of $k_{1}=\omega_0\sqrt{\epsilon_1}/c$. The fundamental 
plasmon mode~(in black) is characterized by a $\kp{\propto}1/R$ dependence. 
Inset: the ratio $\textrm{Re}\;\kp/\textrm{Im}\;\kp$ for the fundamental mode.
 (b) Solid line: Probability of emission into non-plasmon channels, $1-\Gamma_{\footnotesize\textrm{pl}}/(\Gamma'+\Gamma_{\footnotesize\textrm{pl}})$, 
for a nanowire as a function of $R$. Dashed line: probability of emission into 
non-plasmon channels vs. $v$ for a nanotip. Dotted line: 
$1-\tilde{\Gamma}_{\footnotesize\textrm{pl}}(R)/(\Gamma'+\Gamma_{\footnotesize\textrm{pl}})$ 
at final radius $R$ for a nanotip.  Solid points: numerically optimized values 
of 
$1-\tilde{\Gamma}_{\footnotesize\textrm{pl}}(R)/(\Gamma'+\Gamma_{\footnotesize\textrm{pl}})$ 
for a nanotip, obtained via boundary element method simulations. Inset: same 
plot, zoomed in near $R,v=0$. \label{fig:modeconstants}}
\end{figure}

Surface plasmons~\cite{smolyaninov03} are bound, non-radiative electromagnetic 
excitations associated with charge density waves propagating along the surface 
of a conducting object.  For a smooth, cylindrical nanowire, it is convenient to 
write the fields in terms of cylindrical coordinates, 
$\bfE_{j}(\bfr)=\bfE_{j,m}(k_{j\perp}\rho)e^{im\phi}e^{i{\kp}z}$, where $j=1,2$ 
denotes the regions outside and inside the metal, respectively.  The functions 
$\bfE_{j,m}$ are determined by Maxwell's Equations along with appropriate 
boundary conditions and are given in~\cite{stratton41}.  The modes are 
characterized by the longitudinal component of the wave vector $\kp$, which is 
related to the free-space wavevector $k_0=\omega_{0}/c$, dielectric permittivity 
$\epsilon_j$, and imaginary transverse wave vector $\kjperp=i\kappa_{j\perp}$ by 
${\epsilon_j}k_0^2 = \kp^2 -\kappa_{j\perp}^2$.  $\kappa_{j\perp}$ is related to 
the inverse decay length of the field away from the surface.  The dependence of 
$\kp$ for a conducting wire on its radius $R$ is shown in 
Fig.~\ref{fig:modeconstants}(a), for a few lowest order modes.  Throughout this 
paper, all figures and numbers presented use the optical properties of 
silver~($\epsilon_2=-50+0.6i$)~\cite{johnson72} at a vacuum wavelength 
$\lambda_0=1\;\mu$m, and assume a surrounding dielectric $\epsilon_1=2$.

For a conducting 
nanowire~($|\sqrt{\epsilon_i}|k_{0}R{\ll}1,\textrm{Re}\,\epsilon_2<0$), there 
exists one, fundamental TM mode~\cite{takahara97,jahns05} with axial
symmetry~($m=0$), while all higher-order modes are cut off.  For this mode, 
$\kp,\kappa_{j\perp}{\approx}C_{-1}/R$, indicating that the phase velocity 
$v_{ph}{\propto}\omega_{0}R$ is greatly reduced while the transverse mode area 
$A_{\footnotesize{\textrm{eff}}}\propto{R^2}$ is localized to a region on the 
order of the wire size. The proportionality constant $C_{-1}$ depends only on 
$\epsilon_{1,2}$ and is given by
\be \frac{\epsilon_2}{\epsilon_1} \approx 
\frac{2}{(\gamma-\log{2}+\log{C_{-1}})(C_{-1})^2},\label{eq:staticmodecondition} 
\ee
where $\gamma{\approx}0.577$ is Euler's constant.  This sub-wavelength guiding 
of plasmons in metal nanowires has recently been observed in a number of 
experiments~\cite{dickson00,krenn02,ditlbacher05}.

At optical frequencies a conductor has finite losses characterized by 
$\textrm{Im}\;\epsilon_{2}$, resulting in dissipation of the plasmon wave at a 
rate given by $\textrm{Im}\;\kp$. Due to the tighter localization, 
$\textrm{Re}\;\kp/\textrm{Im}\;\kp$ decreases and approaches some non-zero 
value~(${\approx}140$) as $R{\rightarrow}0$, as shown in the inset of 
Fig.~\ref{fig:modeconstants}(a) for the fundamental mode. Physically 
$\textrm{Re}\;\kp/\textrm{Im}\;\kp$ is proportional to the number of plasmon 
wavelengths the plasmon will travel before decaying.

We now consider the emission properties near a nanowire of an oscillating 
dipole, which physically can be formed by a single atom, a defect in a 
solid-state system, or any other system with a dipole-allowed transition.  This 
dipole can generally lose its excitation radiatively by emitting a photon, 
non-radiatively through dissipation of currents induced by the dipole in the 
metal, or into the guided plasmon modes.  It is well-known that the 
corresponding spontaneous emission rates can be obtained via classical 
calculations of the fields~\cite{wylie84}.  For sub-wavelength systems this 
calculation further simplifies because it is sufficient to consider the 
quasi-static~($\bfH=0$) field solutions~\cite{klimov04}, the derivation of which 
we outline here.  Given a point charge source at $\bfrp$ outside the metal, we 
write the static potential outside in terms of the free-space potential and a 
reflected component, $\Phi_{0}(\bfr,\bfrp)+\Phi_{r}(\bfr,\bfrp)$, while 
$\Phi_{2}(\bfr,\bfrp)$ gives the potential inside the wire.  Note that the 
potential due to a dipole $\bfp0$ at $\bfrp$ is easily found from the point 
source potential via 
$\Phi_{dip}(\bfr,\bfrp)=\left(\bfp0\cdot\nabla^{\prime}\right)\Phi(\bfr,\bfrp)$.  
Since $\Phi_{r,2}(\bfr,\bfrp)$ are solutions to the Laplace equation, 
$\nabla^{2}\Phi_{r,2}=0$, we can expand $\Phi_r$ in an appropriate basis,
\be 
\Phi_{r}(\bfr,\bfrp)=\frac{1}{2\pi^2\epsilon_{0}}\sum_{m=0}^{\infty}(2-\delta_{m,0})\cos{m(\phi-\phi')}\int_{0}^{\infty}dh\,\alpha_{m}(h)\cos{h(z-z')}K_{m}(h\rho')I_{m}(h\rho),\label{eq:Phir} 
\ee
where $\alpha_{m}(h)$ are arbitrary coefficients, and $K,I_{m}$ are modified 
Bessel functions.  A similar expansion holds for $\Phi_{2}$, with the 
replacements $\alpha_{m}\rightarrow\beta_m$ and $K_m{\rightarrow}I_m$.  
Expanding $\Phi_0=(4\pi\epsilon_{0}\epsilon_1|\bfr-\bfrp|)^{-1}$ in a similar 
basis, algebraic equations for $\alpha_{m},\beta_{m}$ result by requiring that 
$\Phi_{0}+\Phi_{r}=\Phi_{2}$ at the boundary and that $\epsilon\bfE_{\perp}$ is 
continuous here.  The solution for $\alpha_{m}$ is found to be
\be 
\alpha_{m}(h)=\frac{1}{\epsilon_1}\frac{(\epsilon_2-\epsilon_1)I_{m}^{\prime}(hR)I_{m}(hR)}{\epsilon_{1}I_{m}(hR)K_m^{\prime}(hR)-\epsilon_2K_m(hR)I_{m}^{\prime}(hR)}.\label{eq:alpha} 
\ee

The radiative decay rate is determined by finding the dipole contribution 
${\sim}\frac{\delta{\bf p}\cdot\bfr}{4\pi\epsilon_{1}r^3}$ to 
$\Phi_{dip,r}{\equiv}\left(\bfp0\cdot\nabla'\right)\Phi_{r}$ at large distances 
$r$, which for a nanowire is due to the $m=1$ term in Eq.~(\ref{eq:Phir}). 
Physically $\delta{\bf p}$ corresponds to an induced dipole moment in the 
nanowire, and leads to a modified radiative decay rate 
$\Gamma_{\footnotesize\textrm{rad}}\propto|{\bf p}_{0}+\delta{\bf p}|^2$.  The 
non-radiative and plasmon decay rates can be calculated via 
$\Gamma_{total}{\propto}\textrm{Im}({\bfp0}\cdot{\bf E})$, where ${\bf 
E}=-\nabla\Phi_{dip}(\bfr,\bfrp)|_{\bfr=\bfrp}$ is the total field at the dipole 
location.  In the limit that the distance $d$ between emitter and wire edge 
approaches zero, one finds that ${\bf E}$ diverges due to a substantial 
contribution from the sum over $m$ in Eq.~(\ref{eq:Phir}).  This term is 
proportional to $\textrm{Im}\;\epsilon_2$ and can thus be identified with 
non-radiative decay.  At the same time, Eq.~(\ref{eq:alpha}) exhibits a pole in 
$\alpha_0$ that corresponds to excitation of the fundamental plasmon mode, and 
whose contribution to ${\bf E}$ yields the plasmon decay rate.  For a nanowire, 
the radiative and non-radiative decay rates for a dipole oriented along 
$\hat{\rho}$ are given by~\cite{klimov04}
\bea \Gamma_{\footnotesize\textrm{rad}}/{\Gamma_0} & {\approx} & 
\left|1+\frac{\epsilon_2-\epsilon_1}{\epsilon_2+\epsilon_1}\frac{R^2}{(R+d)^2}\right|^2,
\\ \Gamma_{\footnotesize\textrm{non-rad}}/{\Gamma_0} & \approx & \frac{3}{16k_{0}^{3}d^3\epsilon_1^{3/2}}\textrm{Im}\left(\frac{\epsilon_2-\epsilon_1}{\epsilon_2+\epsilon_1}\right),
\label{eq:Gammaprime} \eea
where $\Gamma_0$ is the decay rate in uniform dielectric 
$\epsilon_{1}$~\cite{note2}, while the plasmon decay rate is~\cite{tobepublished}
\be 
\Gamma_{\footnotesize\textrm{pl}}/\Gamma_0=\alpha_{\footnotesize\textrm{pl}}\frac{K_{1}^{2}\left(\kappa_{1\perp}(R+d)\right)}{(k_0R)^3}. 
\ee
$\alpha_{\footnotesize\textrm{pl}}$ is a complicated expression but depends only 
on $\epsilon_{1,2}$.

The scalings of the various decay rates can be intuitively understood.  Away 
from the plasmon resonance~($\epsilon_1+\epsilon_2{\approx}0$), 
$\Gamma_{\footnotesize\textrm{rad}}$ varies slightly from $\Gamma_0$ due to a 
small change in the radiative density of states near the nanowire, while the 
$1/d^3$ dependence in $\Gamma_{\footnotesize\textrm{non-rad}}$ reflects the 
dissipation of a divergent current induced in the nanowire by the near-field of 
the dipole.  The $1/R^{3}$ scaling for $\Gamma_{\footnotesize\textrm{pl}}$ can 
be understood from Fermi's Golden Rule, 
$\Gamma_{\footnotesize\textrm{pl}}=2{\pi}g^2(\bfr,\omega_0)D(\omega_0)$, where 
$g(\bfr,\omega_0)\propto{1}/\sqrt{A_{\footnotesize{\textrm{eff}}}}\propto{1}/R$ 
is the position-dependent coupling strength between emitter and plasmon modes 
and $D(\omega_0)\propto(d\omega/d\kp)^{-1}\propto(\omega_{0}{R})^{-1}$ is the 
plasmon density of states on the wire.

The position dependence of the decay rates results in an optimal emitter 
distance $d_{o}$ for which the probability of decay into plasmons 
$\Gamma_{\footnotesize\textrm{pl}}/(\Gamma_{\footnotesize\textrm{pl}}+\Gamma^{\prime})$ 
is maximized, where 
$\Gamma^{\prime}=\Gamma_{\footnotesize\textrm{rad}}+\Gamma_{\footnotesize\textrm{non-rad}}$ 
denotes the total ``non-plasmon'' decay rate.  For typical parameters, $d_o$ is 
on the order of several $R$ as $R{\rightarrow}0$, such that the emitter sits 
within the localized plasmon field but is not too close to the wire that 
dissipation become dominant. In Fig.~\ref{fig:modeconstants}(b) we plot the 
probability of emission into non-plasmon channels, 
$1-\Gamma_{\footnotesize\textrm{pl}}/(\Gamma'+\Gamma_{\footnotesize\textrm{pl}})$,
as a function of $R$ when the optimal $d_o$ is chosen. This optimized ``error'' 
rate decreases monotonically as $R{\rightarrow}0$ and approaches a small number 
$\propto\textrm{Im}\;\epsilon/(\textrm{Re}\;\epsilon)^2$ indicating that the 
efficiency of emission into plasmons is ultimately limited by dissipative 
losses. As $R{\rightarrow}0$, one can achieve effective Purcell factors of 
$\Gamma_{\footnotesize\textrm{pl}}/\Gamma'{\approx}5.2{\times}10^2$ for a silver 
nanowire.

\begin{figure}[p]
\begin{center}
\includegraphics[width=9cm]{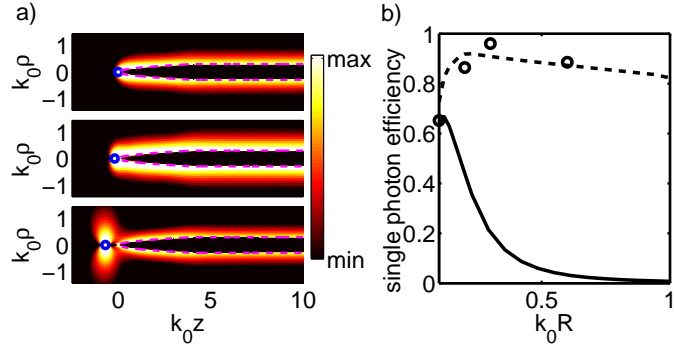}
\end{center}
\caption{(a) Normalized energy flux for an emitter positioned~(from top to 
bottom) at distances $k_{0}d=0.002,0.2,0.7$~(denoted by the blue circles).  The 
nanotip~(whose surface is indicated by the dotted lines) has final radius 
$k_{0}R=0.3$ and curvature parameter $k_{0}v=0.022$. The first plot is mostly 
dark and indicates that the emitter decays primarily non-radiatively.  The 
middle plot demonstrates efficient excitation of guided plasmons at the final 
radius $R$, while the last plot exhibits the typical lobe pattern associated 
with radiative decay. (b) Optimized efficiency of single photon generation vs. 
$R$. We have assumed that coupling to waveguide modes other than the fundamental 
mode is negligible, \textit{i.e.}, the waveguide is effectively in the 
single-mode regime.  Solid line: theoretical efficiency using a nanowire.  
Dotted line: theoretical efficiency using a nanotip. Solid points: efficiency 
based on numerical simulations of emission near a nanotip, combined with 
coupled-mode equations.\label{fig:photonefficiency}}
\end{figure}

The nanowire is a simple system that illustrates the relevant properties of 
dipole emission and plasmon propagation.  One immediately sees, however, that 
the increase in coupling achieved by letting $R{\rightarrow}0$ is accompanied by 
a decrease in the plasmon propagation length, which limits coherent processes of 
interest. Such limits can be circumvented with simple design improvements, which 
we illustrate specifically for the case of a metallic nanotip, assumed to have a 
paraboloidal profile given by $\rho(z)=\sqrt{vz}$~($z>0$).  One expects a 
similar enhancement of plasmon emission due to the nanotip, yet the tip can 
quickly expand to larger radii where losses can be significantly reduced.  As in 
the nanowire case, one can calculate the emission rates based on the 
quasi-static field solution of a dipole near a tip, which is exactly solvable by 
working in parabolic coordinates. For an emitter located on-axis at position 
$z=-|d|<0$ and oriented along $\hat{z}$, we find that~\cite{tobepublished}
\bea \Gamma_{\footnotesize\textrm{rad}}/\Gamma_0 & = & 
\left|1+\left(1+4d/v\right)^{-1}\left(\frac{\epsilon_{2}}{\epsilon_1}-1\right)\right|^2,
\\ \Gamma_{\footnotesize\textrm{non-rad}}/\Gamma_0 & = &
\frac{3}{8\epsilon_{1}^{3/2}(k_{0}d)^3}\textrm{Im}\left(\frac{\epsilon_2-\epsilon_1}{\epsilon_2+\epsilon_1}\right),
\\ \Gamma_{\footnotesize\textrm{pl}}/\Gamma_0 & = &
\alpha_{\footnotesize\textrm{pl}}'\frac{|K_{1}(C_{-1}\sqrt{1+4d/v})|^2}{(k_{0}v)^{3}(1+4d/v)}. 
\eea
Here, $\alpha_{\footnotesize\textrm{pl}}'$ is a constant that again only depends 
on $\epsilon_{1,2}$, and $C_{-1}$ is the solution to 
Eq.~(\ref{eq:staticmodecondition}).  From these decay rates one finds an 
optimized Purcell factor 
$\Gamma_{\footnotesize\textrm{pl}}/\Gamma'{\approx}2.5\times{10^3}$ as 
$v{\rightarrow}0$, as shown in Fig.~\ref{fig:modeconstants}(b), for the same 
material parameters as the nanowire calculation.  In the case of a nanotip, 
however, we are primarily interested in the probability that the plasmon 
propagates up to some final radius $R$.  We estimate this quantity by making an 
eikonal approximation based on the nanowire solution~\cite{stockman04}. In 
particular, we assume that the plasmons are completely emitted into the end of 
the tip~($z=0$) at a rate $\Gamma_{\footnotesize\textrm{pl}}$, while the rate at 
which the plasmon emission successfully propagates to some larger radius $R(z)$ 
is given by 
\be 
\tilde{\Gamma}_{\footnotesize\textrm{pl}}(R)=\Gamma_{\footnotesize\textrm{pl}}\textrm{exp}\left(-2\int_{0}^{z(R)}\textrm{Im}\kp(\rho(z))dz\right). 
\ee
Here $\kp(\rho)$ is the nanowire solution at radius $\rho$.  In 
Fig.~\ref{fig:modeconstants}(b) we plot 
$P_{E}(R)=1-\tilde{\Gamma}_{\footnotesize\textrm{pl}}(R)/(\Gamma'+\Gamma_{\footnotesize\textrm{pl}})$ 
as a function of $R$, optimized over the emitter position and $v$. This quantity 
corresponds to an effective error probability in which the plasmon mode is 
either not excited or fails to successfully propagate to some final radius $R$. 
For $k_{0}R\gtrsim{0.05}$ the tip leads to significant improvement in efficiency 
compared to nanowires of the same $R$.  To check the validity of these 
approximations we have performed numerical~(electrodynamic) simulations of 
dipole emission near a nanotip using boundary element method~\cite{abajo02}, 
with the resulting numerically optimized $P_E(R)$ plotted in 
Fig.~\ref{fig:modeconstants}(b).  It can be seen that the theory agrees well 
with the numerics.  Some typical simulation results are shown in 
Fig.~\ref{fig:photonefficiency}(a).  Here we plot for different emitter 
positions the quantity 
$|\textrm{Re}\;(\bfE\times\bfH^{\ast})|/\Gamma_{\footnotesize\textrm{total}}$, 
which is proportional to the energy flux of the system normalized by the total 
power output of the emitter.  It can clearly be seen that choosing the optimal 
position results in efficient excitation of the plasmons at final radius $R$, 
while other positions can result in primarily non-radiative or radiative decay.

Because of losses, the plasmon modes are not suitable as carriers of information 
over long distances. However, one can evanescently couple the plasmons to 
dielectric waveguide modes, which can form an architecture for a device to 
generate single photons on demand. Noting that the concepts behind single photon 
generation with a single emitter in a cavity have been presented in detail 
elsewhere~\cite{cirac97,michler00,pelton02,mckeever04}, here we illustrate a 
potential novel realization of a single photon device, shown in 
Fig.~\ref{fig:applications}(a). In this architecture, a single, optically 
addressable emitter such as a quantum dot sits atop a nanowire, which 
co-propagates over some length $L_{ex}$ with a nearby dielectric waveguide.  In 
order to maximize the transfer efficiency into the waveguide, the longitudinal 
wavevectors $\kp$ of the plasmon and waveguide should be approximately matched, 
and $L_{ex}$ should be optimized.  In practice, for a given wire radius $R$, the 
matching condition results in some optimization of parameters such as the 
waveguide size.  A similar setup using a nanotip instead of a nanowire is 
illustrated in Fig.~\ref{fig:applications}(b).  Here the nanotip maintains a 
paraboloidal profile up to some final radius $R$, at which point it becomes a 
straight nanowire supporting plasmon modes with well-defined $\kp$ that can be 
easily be mode-matched with the waveguide.

The out-coupling and single photon efficiencies can be calculated using standard 
mode-coupled equations based on Lorentz reciprocity~\cite{barclay03}.  For 
simplicity, we take the dielectric waveguide to be a cylindrical optical fiber, 
whose modes can be calculated analytically~\cite{stratton41}, and set the 
surrounding and fiber core permittivities to be $\epsilon_1=2,\epsilon_c=13$.  
In Fig.~\ref{fig:photonefficiency}(b), we plot the optimized efficiency $P$ for 
single photon generation as a function of $R$, for both the nanowire and 
nanotip, based on the decay probabilities obtained above and coupled-mode 
theory.  We also include the predicted efficiencies using the boundary element 
method simulations combined with coupled-mode theory.  These calculations take 
fully into account all imperfections, including metal losses and imperfect 
waveguide coupling.  We observe that there is some optimal $R$ where $P$ is 
maximized, which corresponds to a balance between achieving large coupling 
between emitter and nano-structure, and ensuring that the plasmon/guide coupling 
exceeds the enhanced losses at small $R$.  We find that optimal single photon 
efficiencies exceeding $95\%$ are achievable in such a system.

Such an architecture for quantum communication based on plasmonic devices has 
several important features. First, unlike typical methods of cavity QED, the 
plasmon excitation covers a broad bandwidth and requires no special tuning to 
achieve resonance.  The operation speeds can also be quite high because of the 
sub-wavelength mode volumes associated with the plasmons. Finally, we note that 
rapid advances in recent years in fabrication techniques for 
nanowires~\cite{xia02,schultz02}, nanotips~\cite{libioulle95}, and 
sub-wavelength dielectric waveguides~\cite{tong03,vlasov04} puts such a system 
in experimental reach.  

The authors thank Atac Imamoglu for useful discussions.  This work was supported 
by the ARO-MURI, ARDA, NSF, the Sloan and Packard Foundations, and by the Danish 
Natural Science Research Council.



\end{document}